# General Solution to the Mixing Problem: Application to Medical Research and Diagnostics

Neil Zhao[a,*]

[a]*Thomas Jefferson University, Philadelphia, Pennsylvania, USA*

The mixing problem is classically encountered in the study of differential equations applied to fluid dynamics. An understanding of fluid movement under constraints is particularly important in the field of medicine as many therapeutics and biologic molecules are dissolved in bodily fluids. Many areas of biomedical research and diagnostics also rely on fluid sampling to obtain accurate measurements of biologic markers. We present in this manuscript the general solution to the mixing problem in the context of studying physiological phenomena based on the movement of fluid acting as a carrier for medically relevant molecules/solutes. We also expanded the general solution to become more compatible with areas of biomedical research and diagnostics that seek to characterize bodily fluids located in areas that are difficult to sample.

*E-mail of corresponding author: neil.zhao@students.jefferson.edu

# 1 Introduction

The mixing problem is a classic question in the study of differential equations. The scenario is usually presented as thus: A container of liquid holding some concentration of solute is connected to an inflow and an outflow. Solvent inflows at a given rate, and the contents of the container outflow at a given rate. The task is to express the concentration of the solute as a function of time. While special cases such as constant and equal inflow and outflow can be solved analytically, a general solution with the least number of assumptions can only be solved numerically.

The setup of the mixing problem can be superimposed on many biological phenomena. Physiological processes like blood flow through an organ or the flow of lymphatic fluid through lymph nodes can be viewed as inflow and outflow from a defined reservoir. The solute can thus be viewed as biological molecules like nutrients, cytokines, molecular signals, or perhaps therapeutics. Mathematical models of lymphatic flow have been constructed to characterize the movement of fluid through lymph nodes [1–3]. Blood flow through the kidney has also been mathematically modeled to simulate glomerular filtration and study the kidney's mechanisms of self-regulation [4]. The basic principles of fluid dynamics can be applied to the movement of blood in blood vessels to generate simple yet accurate models of the circulatory system [5].

We present in this manuscript the general solution to the mixing problem in the context of applying it to study physiological processes. We also expanded the classic setup of the mixing problem to broaden the applicability of the general solution to medical research and diagnostics.

## 2 General solution to the mixing problem

We begin with a model of fluid flow into and out of a container, known as the inflow reservoir (Fig. 1). A solute is dissolved in the inflow reservoir, while the inflow itself contains no solute; solute is also found in the outflow.

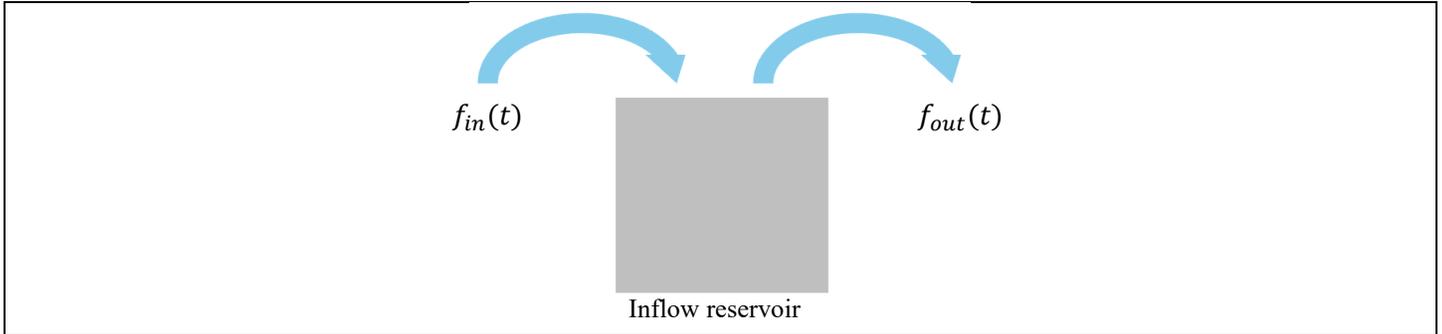

Figure 1: A schematic of fluid inflow and outflow in the inflow reservoir. Inflow reservoir contains solute with concentration $\rho(t)$.

---
**I. Definitions [units]**

$\rho_o = \text{initial concentration } \left[\frac{mass}{volume}\right]$

$m(t) = mass, function\ of\ time$

$\rho(t) = concentration, function\ of\ time$

$V_o = initial\ volume$

$V(t) = volume, function\ of\ time$

$f_{in}(t) = inflow, function\ of\ time\ \left[\frac{volume}{time}\right]$

$f_{out}(t) = \text{outflow}, function\ of\ time$

---

Change in mass of solute can be written in two formats:

$$\Delta m(t) = \Delta\big(\rho(t) * V(t)\big) = V(t) * \Delta\rho(t) + \rho(t) * \Delta V(t) \qquad (1)$$

$$\Delta m(t) = -\rho(t) * f_{out}(t) * \Delta t \qquad (2)$$

Volume of the inflow reservoir can be written as follows:

$$V(t) = V_o + \int_0^t \big(f_{in}(t) - f_{out}(t)\big)\, dt \qquad (3)$$

Change in volume is thus written:

$$\Delta V(t) = \big(f_{in}(t) - f_{out}(t)\big) * \Delta t \qquad (4)$$

Equate the two equations for change in mass:

$$-\rho(t) * f_{out}(t) * \Delta t = V(t) * \Delta\rho(t) + \rho(t) * \Delta V(t) \qquad (1)\ (2)$$

Substitute the $V(t), \Delta V(t)$ on the right side:

$$= \left( V_o + \int_0^t \left( f_{in}(t) - f_{out}(t) \right) dt \right) * \Delta\rho(t) + \rho(t) * \left( f_{in}(t) - f_{out}(t) \right) * \Delta t \tag{5}$$

Simplify and rearrange:

$$-\rho(t) * f_{in}(t) * \Delta t = \left( V_o + \int_0^t \left( f_{in}(t) - f_{out}(t) \right) dt \right) * \Delta\rho(t) \tag{6}$$

$$\frac{-f_{in}(t) * \Delta t}{V_o + \int_0^t \left( f_{in}(t) - f_{out}(t) \right) dt} = \frac{\Delta\rho(t)}{\rho(t)} \tag{7}$$

Integrate:

$$\int_0^t \frac{-f_{in}(t)\, dt}{V_o + \left[ \int_0^t \left( f_{in}(t) - f_{out}(t) \right) dt \right]} = \int_{\rho_o}^{\rho} \frac{d\rho(t)}{\rho(t)} \tag{8}$$

Integrating (8) provides the general solution of concentration of solute in the inflow reservoir as a function of time. An assumption that carries through to this step is that the initial volume and concentration of the inflow reservoir is not equal to zero.

$$\boxed{\ln\left(\frac{\rho(t)}{\rho_o}\right) = \int_0^t \frac{-f_{in}(t)\, dt}{V_o + \left[ \int_0^t \left( f_{in}(t) - f_{out}(t) \right) dt \right]}} \tag{9}$$

A special condition to consider is when inflow = outflow, $f_{in}(t) = f_{out}(t)$. Such a condition is relevant in situations where the inflow reservoir is quickly filled but cannot expand due to physical constraints.

$$\ln\left(\frac{\rho(t)}{\rho_o}\right) = \int_0^t \frac{-f_{in}(t)\, dt}{V_o + \left[ \int_0^t \left( f_{in}(t) - f_{out}(t) \right) dt \right]} \tag{9}$$

$$\ln\left(\frac{\rho(t)}{\rho_o}\right) = \int_0^t \frac{-f_{in}(t)\, dt}{V_o + 0} \tag{10}$$

$$\rho(t) = \rho_o * e^{\frac{-1}{V_o} \int_0^t f_{in}(t)\, dt} \tag{11}$$

Case 1: Assume constant inflow and outflow, $C$

$$\rho(t) = \rho_o * e^{\frac{-C*t}{V_o}} \tag{12} \quad \text{(Fig. 3A, C)}$$

When the inflow and outflow are equal and constant, the solute concentration in the inflow reservoir decreases exponentially with time. This is further made evident in (Fig. 3C), where the logarithmic transformation of the inflow reservoir results in a line.

Case 2: Assume exponential inflow and outflow, $A * e^{-k*t}$

$$\rho(t) = \rho_o * e^{\frac{-1}{V_o} \int_0^t A*e^{-k*t} dt} \qquad (13)$$

$$\rho(t) = \rho_o * e^{\frac{A}{V_o*k}*(e^{-k*t}-1)} \qquad (14) \text{ (Fig. 3B, D)}$$

In this case, the change in solute concentration in the inflow reservoir is not strictly exponential; the logarithmic transformation does not result in a line (Fig. 3D). Furthermore, an assumption so far is that the inflow has no solute concentration. This does not hold true in closed systems where the outflow is eventually circulated back into the inflow. We accounted for a closed system with non-zero inflow solute concentration in the supplement. In the remaining analysis in this section, we continued to assume that the inflow solute concentration is negligible.

We now supplement our model by adding an outflow reservoir that collects the outflow with the dissolved solute.

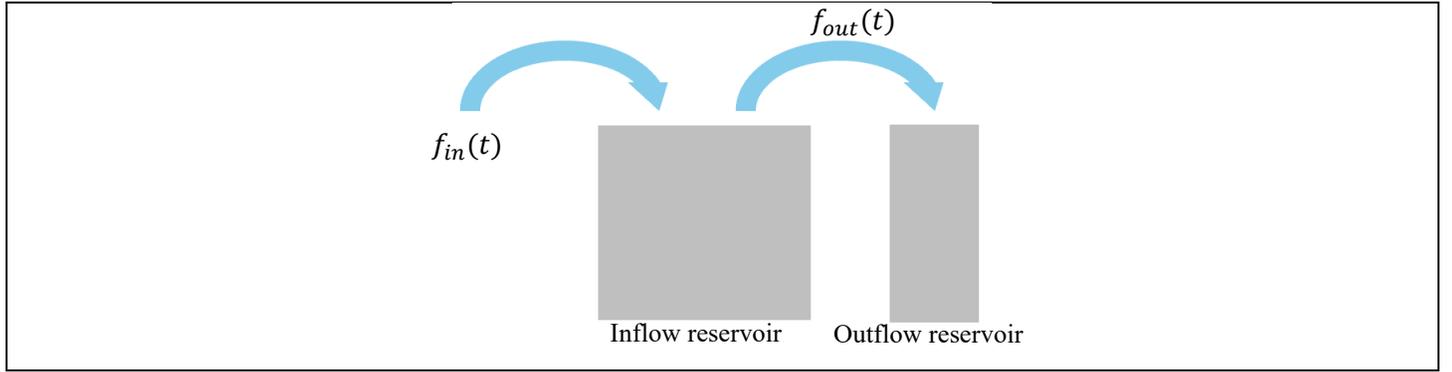

Figure 2: A schematic of fluid inflow and outflow in the inflow reservoir and outflow reservoir. Inflow reservoir contains solute with concentration $\rho(t)$. Outflow reservoir contains solute with concentration $\rho_{out}(t)$.

## II. Definitions

$\rho_{out}(t) = concentration\ in\ total\ outflow, function\ of\ time$
$m_{out}(t) = mass\ of\ total\ outflow, function\ of\ time$
$V_{out}(t) = volume\ of\ total\ outflow, function\ of\ time$

Change in mass of solute in outflow reservoir can be written as follows:

$$\Delta m_{out}(t) = \rho(t) * f_{out}(t) * \Delta t \qquad (15)$$

$$\int_{m_o}^m dm_{out}(t) = \int_0^t \rho(t) * f_{out}(t)\, dt \qquad (16)$$

Change in volume of outflow reservoir can be written as follows:

$$\Delta V_{out}(t) = f_{out}(t) * \Delta t \qquad (17)$$

$$\int_{V_o}^V dV_{out}(t) = \int_0^t f_{out}(t)\, dt \qquad (18)$$

Dividing mass by volume provides the general solution of the solute concentration in the outflow reservoir as a function of time. The assumption that the initial volume and concentration of the inflow reservoir is not equal to zero carries through to here.

$$\boxed{\rho_{out}(t) = \frac{m_{out}(t)}{V_{out}(t)} = \frac{\int_0^t \rho(t) * f_{out}(t)\, dt}{\int_0^t f_{out}(t)\, dt}} \qquad (16)\ (18)$$

We again consider the special condition when inflow = outflow, $f_{in}(t) = f_{out}(t)$.

$$\rho(t) = \rho_o * e^{\frac{-1}{V_o} \int_0^t f_{in}(t)\, dt} \qquad (11)$$

$$\rho_{out}(t) = \frac{\int_0^t \left[ \rho_o * e^{\frac{-1}{V_o} \int_0^t f_{in}(t)\, dt} * f_{out}(t) \right] dt}{\int_0^t f_{out}(t)\, dt} \qquad (19)$$

Case 1: Assume constant inflow and outflow, $C$

$$\rho_{out}(t) = \frac{\rho_o \int_0^t e^{\frac{-C*t}{V_o}}\, dt}{t} \qquad (20)$$

$$\rho_{out}(t) = \frac{\rho_o * V_o}{t * C} * \left( 1 - e^{\frac{-C*t}{V_o}} \right) \qquad (21)\ (\text{Fig. 3A})$$

The outflow reservoir concentration remains higher than the inflow reservoir at all timepoints (Fig. 3A). Additionally, the change in concentration is not strictly exponential. Logarithmic transformation does not result in a line (Fig. 3C).

Case 2: Assume exponential inflow and outflow, $A * e^{-k*t}$

$$\rho_{out}(t) = \frac{\int_0^t \left[ \rho_o * e^{\frac{-1}{V_o} \int_0^t A*e^{-k*t}\, dt} * A * e^{-k*t} \right] dt}{\int_0^t A * e^{-k*t}\, dt} \qquad (22)$$

$$\int_0^t \left[ \rho_o * e^{\frac{-1}{V_o} \int_0^t A*e^{-k*t}\, dt} * A * e^{-k*t} \right] dt = \int_0^t \left[ A * \rho_o * e^{\frac{-A}{V_o * k}*(1-e^{-k*t})-k*t} \right] dt \qquad (23)$$

$$\int_0^t A * e^{-k*t}\, dt = \frac{A}{k} * (1 - e^{-k*t}) \qquad (24)$$

$$\rho_{out}(t) = \frac{\rho_o * k \int_0^t e^{\frac{-A}{V_o * k}*(1-e^{-k*t})-k*t}\, dt}{1 - e^{-k*t}} \qquad (23)\ (24)\ (\text{Fig. 3B})$$

Like constant inflow and outflow, the outflow reservoir concentration remains higher than the inflow reservoir at all timepoints. The change in concentration is also not strictly exponential, as given by the logarithmic transformation not being a line (Fig. 3D).

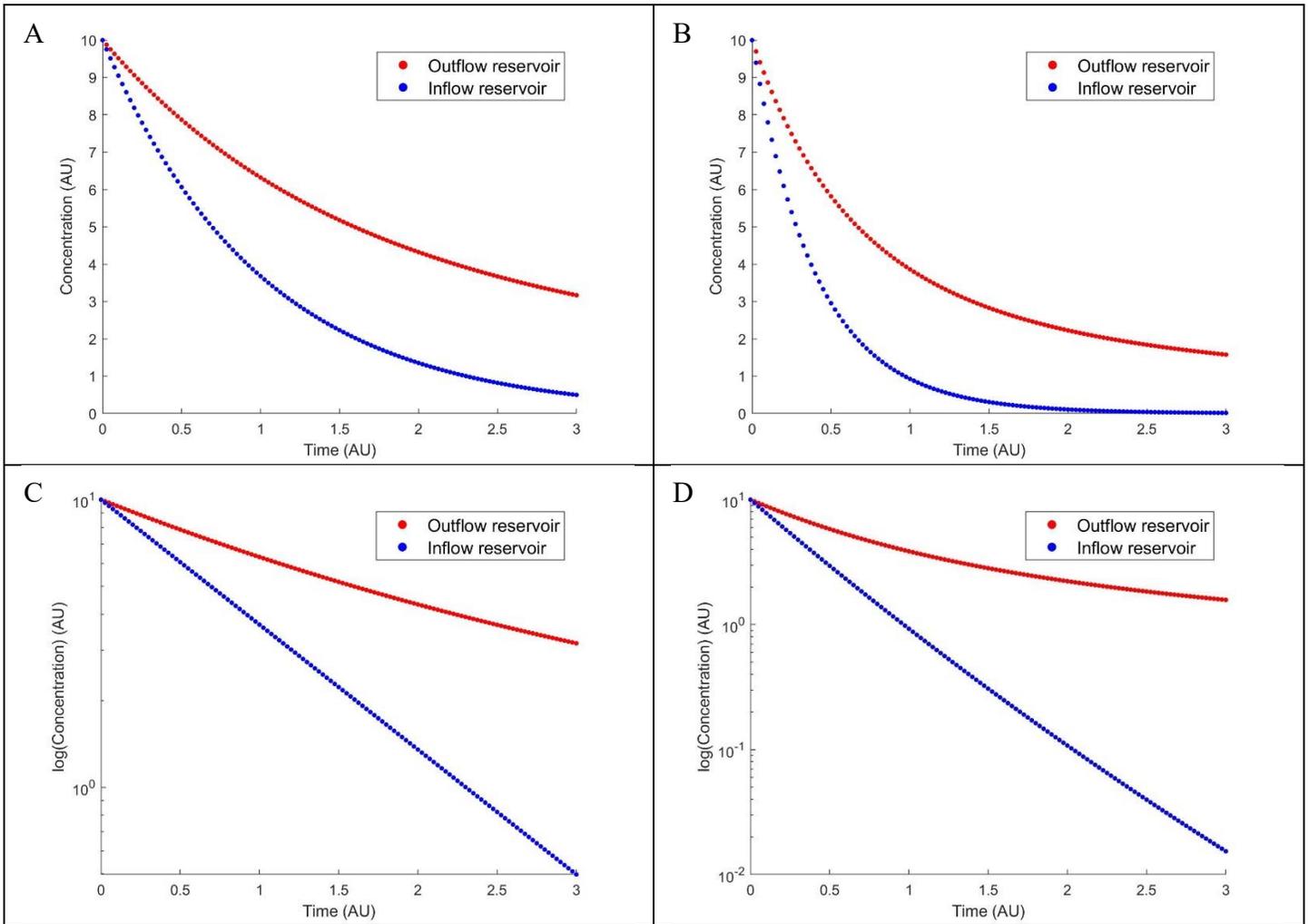

Figure 3: Solute concentrations in the inflow and outflow reservoirs. (A) Linear and (C) Logarithmic scale of Inflow = Outflow = constant, $C$. (B) Linear and (D) Logarithmic scale of Inflow = Outflow = exponential, $A * e^{-k*t}$. Arbitrary units (AU) are indicated along the x-axis and y-axis. Numerical solutions were calculated and graphed using MATLAB 2019a.

# 3 Conclusion

The general solution to the mixing problem can be numerically solved and remains valid with the assumption that the initial concentration of the inflow reservoir is not equal to zero. One is also able to adjust the inflow and outflow based on the desired level of complexity for the system: $f_{in}(t) = \sum_i f_{in,i}$ and $f_{out}(t) = \sum_i f_{out,i}$, where the net inflow and outflow can be represented as the sums of different functions of time. The simplicity and versatility of the general solution lead to its application in modeling biological phenomena. Human physiology and medical science present many situations where a solute (therapeutics, nutrients, cytokines, molecular signals) is dissolved in an enclosed space with multiple external conduits. The movement of bodily fluid (blood, lymphatic fluid, interstitial fluid, wound fluid, cerebral spinal fluid) through these channels leads to changes in solute concentration, which can have implications for their intended physiological effects. The same principles and assumptions used in this manuscript in presenting the general solution to the mixing problem can be used to provide quantitative estimates of these changing solute concentrations.

Our supplementation of the mixing problem to include an outflow reservoir extends the applicability of our results to biomedical research and diagnostics. Direct sampling of certain enclosed spaces within the body is unfeasible or entails a high risk of iatrogenic harm. Examples include locations near the spinal cord and within the brain. In such situations, a proxy location is often chosen for sampling, despite the higher variability and error in these measurements. Our extension of the general solution to include an outflow reservoir can be used to provide researchers and physicians with an estimate of that variability and error.

A potential concern is that the background bodily concentration of the solute will increase over an extended time. Consequently, the inflow will begin to carry dissolved solute into the inflow reservoir. This is an important consideration that must be accounted for when 1) the concentration of the solute in the reservoir is on the same order of magnitude as the general background bodily concentration over a long period of time, and 2) the analysis is extended over a time when the concentration of the solute in the reservoir decreases to the general background concentration. We showed in the supplement that the effects of background bodily concentration are small when the initial solute concentration in the inflow reservoir is ≥ 20X the background concentration. This requirement is readily achievable in many circumstances due to the high water content of the adult body habitus. Therefore, any contributions from solute present in the inflow in closed systems under conditions that satisfy those stated above can generally be neglected.

# 4 Supplement

Further generalization of the mixing problem to include inflow with non-negligible solute concentration given by $\delta(t)$.

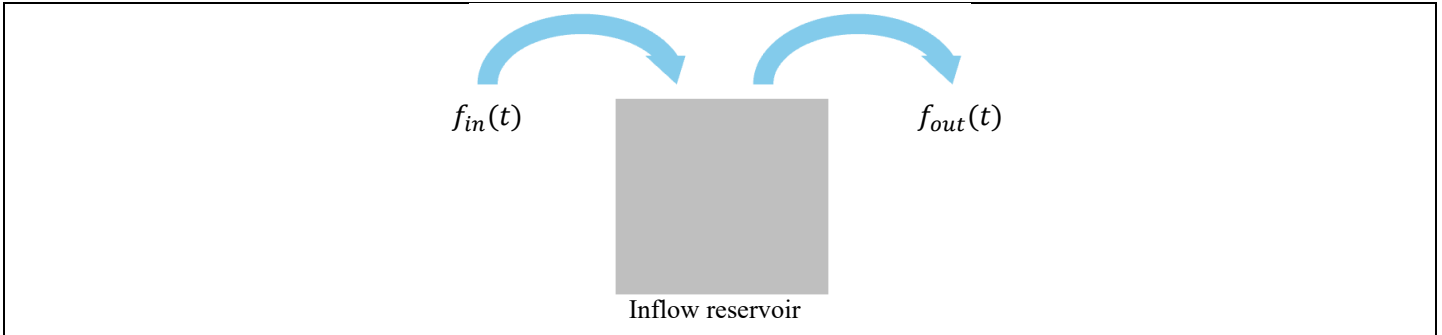

Supplementary Figure 1: A schematic of fluid inflow and outflow in the inflow reservoir. Inflow reservoir contains solute with concentration $\rho(t)$. Inflow contains solute with concentration $\delta(t)$.

**I. Definitions [units]**

$$\rho_o = \text{initial concentration} \left[\frac{mass}{volume}\right]$$
$$m(t) = \text{mass, function of time}$$
$$\rho(t) = \text{concentration, function of time}$$
$$\delta(t) = \text{inflow solute concenttration, function of time}$$
$$V_o = \text{initial volume}$$
$$V(t) = \text{volume, function of time}$$
$$f_{in}(t) = \text{inflow, function of time} \left[\frac{volume}{time}\right]$$
$$f_{out}(t) = \text{outflow, function of time}$$

Change in mass of solute can be written in two formats:

$$\Delta m(t) = \Delta\big(\rho(t) * V(t)\big) = V(t) * \Delta\rho(t) + \rho(t) * \Delta V(t) \tag{S1}$$

$$\Delta m(t) = \delta(t) * f_{in}(t) * \Delta t - \rho(t) * f_{out}(t) * \Delta t \tag{S2}$$

Volume of the inflow reservoir can be written as follows:

$$V(t) = V_o + \int_0^t \big(f_{in}(t) - f_{out}(t)\big)\, dt \tag{S3}$$

Change in volume is thus written:

$$\Delta V(t) = \big(f_{in}(t) - f_{out}(t)\big) * \Delta t \tag{S4}$$

Equate the two equations for change in mass:

$$\delta(t) * f_{in}(t) * \Delta t - \rho(t) * f_{out}(t) * \Delta t = V(t) * \Delta\rho(t) + \rho(t) * \Delta V(t) \tag{S1) (S2}$$

Substitute the $V(t), \Delta V(t)$ on the right side:

$$= \left( V_o + \int_0^t (f_{in}(t) - f_{out}(t)) \, dt \right) * \Delta \rho(t) + \rho(t) * (f_{in}(t) - f_{out}(t)) * \Delta t \tag{S5}$$

Simplify and rearrange:

$$(\delta(t) - \rho(t)) * f_{in}(t) * \Delta t = \left( V_o + \int_0^t (f_{in}(t) - f_{out}(t)) \, dt \right) * \Delta \rho(t) \tag{S6}$$

$$\frac{(\delta(t) - \rho(t)) * f_{in}(t)}{V_o + \int_0^t (f_{in}(t) - f_{out}(t)) \, dt} = \frac{\Delta \rho(t)}{\Delta t} \tag{S7}$$

Take the limit as $\Delta t \to 0$:

$$\boxed{\frac{(\delta(t) - \rho(t)) * f_{in}(t)}{V_o + \int_0^t (f_{in}(t) - f_{out}(t)) \, dt} = \frac{d\rho(t)}{dt}} \tag{S8}$$

Equation (S8) cannot be solved analytically. We will consider some special conditions to solve (S8) numerically. We will again assume an exponential inflow = outflow, $f_{in}(t) = f_{out}(t) = A * e^{-k*t}$. We will also assume that the inflow solute concentration begins at 0 and asymptotically approaches a dilution of $\rho_o$ by a factor of $L$, $\delta(t) = \frac{\rho_o}{L} * (1 - e^{-r*t})$.

Substitution of equations into (S8):

$$\frac{\left(\frac{\rho_o}{L} * (1 - e^{-r*t}) - \rho(t)\right) * A * e^{-k*t}}{V_o + 0} = \frac{d\rho(t)}{dt} \tag{S9}$$

Simplify and rearrange:

$$\frac{A}{V_o} * \left( \frac{\rho_o}{L} * (1 - e^{-r*t}) - \rho(t) \right) * e^{-k*t} = \frac{d\rho(t)}{dt} \tag{S10}$$

Solution when inflow solute concentration $\delta(t) = 0$

$$\rho(t) = \rho_o * e^{\frac{A}{V_o * k} * (e^{-k*t} - 1)} \tag{11}$$

The numerical solution of (S10) with the dilution factor $L$ ranging from 100 to 5 is graphed in Supplementary Figure 2 along with the base case solution assuming $\delta(t) = 0$ (11). Deviation from the base case solution increased at higher time points with lower dilution factors.

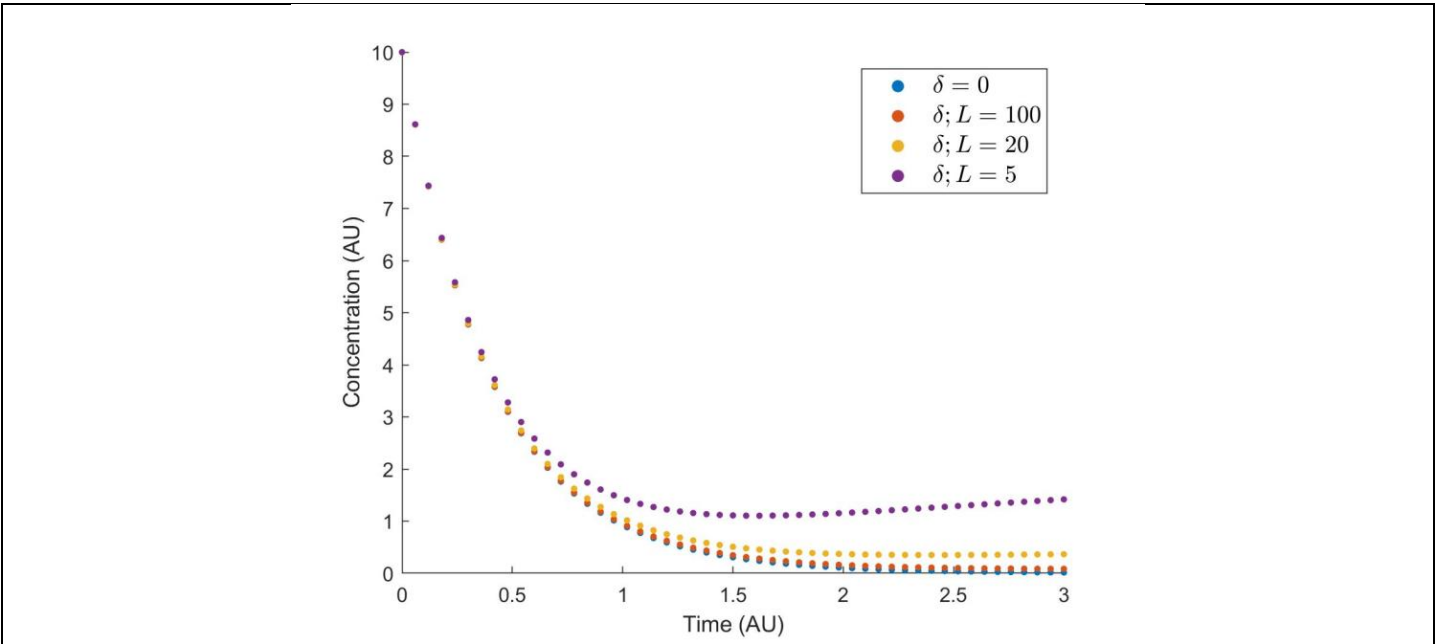

Supplementary Figure 2: Solutions to the mixing problem with non-zero inflow solute concentration, $\delta(t) = \frac{\rho_o}{L} * (1 - e^{-r*t})$ compared with the solution assuming zero inflow solute concentration, $\delta(t) = 0$. Arbitrary units (AU) are indicated along the x-axis and y-axis. Numerical solutions were calculated and graphed using Simulink in MATLAB 2019a.

# References

[1] Jayathungage Don T D, Safaei S, Maso Talou G D, Russell P S, Phillips A R J and Reynolds H M 2023 Computational fluid dynamic modeling of the lymphatic system: a review of existing models and future directions. *Biomech Model Mechanobiol*

[2] Giantesio G, Girelli A and Musesti A 2021 A model of the pulsatile fluid flow in the lymph node *Mech. Res. Commun.* **116** 103743

[3] Giantesio G, Girelli A and Musesti A 2022 A mathematical description of the flow in a spherical lymph node. *Bull Math Biol* **84** 142

[4] Sgouralis I and Layton A T 2015 Mathematical modeling of renal hemodynamics in physiology and pathophysiology. *Math Biosci* **264** 8–20

[5] Khalid A K, Othman Z S and M Shafee C M N 2021 A review of mathematical modelling of blood flow in human circulatory system *J. Phys.: Conf. Ser.* **1988** 012010